
\input phyzzx
\vskip 1cm
\centerline{{\bf The $\mu$ Problem and the Invisible Axion
}\foot{Talk presented at XXVI Int. Conf. High Energy Physics,
Dallas, Texas, U.S.A., August 6--12, 1992.}}

\vskip 3cm
\centerline{Jihn E. Kim}

\vskip 0.5cm
\centerline{\it Center for Theoretical Physics and Department
of Physics}
\centerline{\it Seoul National University}
\centerline{\it Seoul 151-742, Korea}

\vskip 4cm
\centerline{\sl Abstract}

{The $\mu$ term in the supersymmetric standard model is known  to
be nonzero.  Supersymmetry breaking at the intermediate scale
may provide the needed $\mu$ term and the invisible axion.  A
possible solution of this $\mu$ problem in superstring models
is also discussed.}

\endpage

\noindent 1. {THE $\mu$ PROBLEM}

Supersymmetry has been introduced to solve the scalar mass
problem or the gauge hierarchy problem.
The minimal supersymmetric standard model (MSSM) contains two
Higgs doublets
$$ H_1 (Y=-\half),\ \ H_2 (Y=\half).\eqno (1)$$
Due to the above hypercharge assignment, $H_1$ couples to $d_L^c$
and $H_2$ couples to $u_L^c$.  This type of couplings that $Q=
-1/3$ quarks get masses from one type of Higgs doublets and $Q=
2/3$  quarks get masses from the other Higgs field
hints a possibility of a
Peccei-Quinn (PQ) symmetry.   Introducing singlet superfields,
one can introduce Higgs quartic couplings.  However, the quartic
couplings are absent in MSSM due to the absence of singlet
superfields.  The Higgs couplings in
supersymmetric theory can be conveniently given in terms of
superpotential.  The superpotential
$W$ in MSSM can contain a term
$$W_{\mu}=\mu H_1H_2 \eqno (2)$$
where $\mu$ is a free parameter.
If $\mu=0$, there results a spontaneously
broken global PQ symmetry and an axion with $m_a\sim
0.1$ MeV, which is  phenomenologically  ruled  out.   Also,  the
parameter space allowed by LEP experiments favors $\mu\sim
100$ GeV.  In supergravity models, the electroweak scale
of 100 GeV arises as soft supersymmetry breaking terms through
the gravitino mass, $m_{3/2}\sim M_I^2/M_{P}$, where $M_P$ is the

Planck mass.  The parameter for the electroweak symmetry breaking
is provided if $M_I\sim 10^{11}$ GeV.  The soft supersymmetry
breaking terms are of order of the gravitino mass
$$Am_{3/2}W_\mu +{\rm h.c.}+Bm_{3/2}^2\sum_{i}\phi^*\phi+\cdots
 \eqno (3)$$
where $A$ and $B$ are dimensionless numbers of $O(1)$.
But the supersymmetric $\mu$
term in the superpotential is not of this kind of a soft term
and must be put in by hand, which is the so-called $\mu$ {\it
problem}.  Eqs. (2) and (3) shows that PQ symmetry is broken by
the presence of the $\mu$ term.
The $\mu$ problem is a fine tuning problem of why $\mu$ must be
so small compared to a high energy scale.   The best motivation
for introducing supersymmetry has been to solve the gauge
hierarchy problem.    If we put in the $\mu$ term by hand, we
lose the original motivation of introducing the supersymmetry
in particle physics.    Since
$\mu\rightarrow 0$ gives a PQ symmetry, any radiative generation
scheme of $\mu$ must start from a PQ symmetry in the beginning.
Thus, studying the $\mu$ problem from the PQ symmetry can be
applied to most models generating an electroweak scale $\mu$.

\noindent 2. {COMMON SCALE}

The hidden sector scale in supergravity models and the
invisible axion scale fall in the common region; it is  desirable
to have them tied together.
Because the $\mu$ term signals the breaking of the Peccei-Quinn
symmetry, it is logical to understand it from the symmetry
principle [1].  Thus we introduce a PQ symmetry in supergravity
models.  Through the supergravity interaction,
$$W\sim {1\over M_P}S_1S_2H_1H_2\eqno (4)$$
can be generated where $S_1$ and $S_2$ are $SU(3)\times
SU(2)\times U(1)$ singlet superfields carrying nonvanishing
net Peccei-Quinn charge.
The scale of the Peccei-Quinn symmetry breaking by vacuum
expectation values of scalar components of $S_1$ and $S_2$
at $10^{10}\sim 10^{13}$ GeV can lead to an electroweak
scale $\mu$ term,
$$\mu \simeq {\langle\bar s_1s_2\rangle\over M_P} \sim
{(10^{10-11}\ {\rm GeV})^2\over M_P} \sim  1000\ {\rm GeV}.
\eqno (5) $$
Supergravity interactions
can    generate   the   needed   global    symmetry    preserving
nonrenormalizable superpotential Eq. (4).

\endpage
\noindent 3. {PQ SYMMETRY WITH ANOMALOUS U(1)}

It is well known that superstring models in ten dimensions
do not allow any global symmetry.  In addition,
there exists the scale problem of the model-independent axion
[2] in
superstring models, $F_a\sim 10^{15-16}$ GeV [3].   This
model-independent axion realizes a global symmetry

nonlinearly.  The scale problem of the model-independent axion
was a serious problem in superstring models [3].  However,
this was a story before the discovery of anomalous $U(1)$
gauge symmetry in some compactification schemes [4].  In other
words, if a compactified theory does not
have any gauge anomaly, the strong CP problem cannot be solved
by an invisible axion because of no global symmetry
available  or too large axion decay constant.   However,  some
compactification schemes allow anomalous $U(1)$ gauge symmetry,
which is not inconsistent because of the model independent-axion
[5].
The model-independent axion becomes the longitudinal degree of
the anomalous $U(1)$ gauge boson.  Then the theory becomes
consistent below the $U(1)$ gauge boson mass scale, and there
survives a global symmetry $U(1)_{PQ}$.  This is due to the
't Hooft mechanism [6] that a global symmetry survives if a
gauge symmetry and a global symmetry is broken by the VEV of
a single Higgs field rendering the gauge boson a mass.  In
the present case, the mechanism works as follows.
Let the gauge boson of the anomalous $U(1)$ be $A_\mu$ and
the model-independent axion $a_{MI}$.  The corresponding currents
are not divergenceless,
$$\eqalign{\partial^\mu J_\mu^A=\Gamma_A {1\over 32\pi^2}
F\tilde F\cr
\partial^\mu J_\mu^a=\Gamma_a{1\over 32\pi^2}F\tilde F
}\eqno (6)$$
where $F\tilde F$ is an abbreviation for $2 {\rm Tr}F_{\mu\nu}
\tilde F^{\mu\nu}$.
Therefore, we can write an effective interaction in the form,
$${\cal L}_{int}=A^\mu J_\mu^A-\Gamma_a{a_{MI}\over f}{1\over 32
\pi^2}F\tilde F.\eqno (7)$$
The transformations of the
anomalous $U(1)_A$ and global $U(1)_a$ are
$$\eqalign{A_\mu\ &\rightarrow A_\mu +\partial_\mu
\Lambda (x)\cr a_{MI} &\rightarrow a_{MI}+f\theta.} \eqno (8)$$
Under these transformations, the shift of ${\cal L}_{int}$ is
$$\delta {\cal L}_{int}=(-\Lambda \Gamma_A-\theta\Gamma_a){1\over
32\pi^2}F\tilde F.\eqno (9) $$
Choosing
$$\theta=-{\Gamma_A\over \Gamma_a}\Lambda (x),\eqno (10) $$
${\cal L}_{int}$ remains invariant.  The
the   nonlinearly   realized   global
transformation is identified as  the gauge transfomation, and
$a_{MI}$ becomes the longitudinal degree of the original
anomalous gauge boson.  The mass of the gauge boson becomes
$$M_A^2=\left({\Gamma_A\over \Gamma_a}f\right)^2.\eqno (11) $$
Below the scale $M_A$, we can define a anomalous global
current
$$J_\mu^{gl}\equiv {\Gamma_A\over \Gamma_a}J^A_\mu +J^a_\mu
\eqno (12.a)$$
and  the  other  orthogonal  combination  is  anomaly  free   and
corresponds to the broken gauge symmetry
$$J_\mu^{ga}\equiv J^A_\mu-{\Gamma_A\over \Gamma_a}J^a_\mu .
 \eqno (12.b)$$
Below the scale $M_A$, a global symmetry survives.   Thus,

in superstring models with anomalous $U(1)$ there exists a
possibility of solving the strong CP problem by an invisible
axion  with  its decay constant at $F_a\sim  10^{12}$  GeV.
Therefore, if there exists only one $\theta$ to remove below the
anomalouss  gauge boson mass, then we solve the scale problem  of
the model independent axion.  The $U(1)_{PQ}$ symmetry is broken
when $\langle S_1S_2\rangle$ get VEV around $10^{12}$ GeV.

\noindent 4. {SUPERSTRINGS AND THE $\mu$ PARAMETER}

For
supersymmetry  breaking  by gaugino condensation,  one  needs  an
extra  confining gauge group at $\sim 10^{13}$ GeV [7]
and there exists one more $\theta$.  Thus the global symmetry
from the anomalous $U(1)$ gauge symmetry is not enough to remove
two $\theta$'s.
For example, an orbifold
construction with the shift and Wilson lines [8]
$$\eqalign{ v&= (11112000)(20000000)\cr
&= (00000002)(01100000)\cr
&= (11121011)(11000000)\cr
&= (00000200)(00011112)}\eqno (13)$$
gives the gauge group
$$SU(3)\otimes SU(2)\otimes U(1)^5\otimes [SU(5)\otimes U(1)^4
]^\prime .\eqno (14) $$
Three generations of doublet quarks in addition to other light
fields arise from the untwisted sector
$$3(3,2,1)+3(\bar 3,1,1)+3(1,2,1)+3(1,1,5^\prime).\eqno (15.a)$$
The light fermions from the twisted sector are
$$9(3,1,1)+12(\bar 3,1,1)+30(1,2,1)+3(1,1,\bar 5^\prime)
+{\rm singlets}.\eqno (15.b) $$
There is no triangle anomalies except for one anomalous
gauge $U(1)_X$.

The explicit calculation of the
sums of $X$ charges for color triplets, $SU(2)$ doublets, and
hidden color quintets are nonvanishing and equal,
$$ \partial^\mu J^X_\mu = c\left( F\tilde F+W\tilde W+F^\prime
\tilde F^\prime +\cdots \right)\eqno (16)$$
where   $\cdots$   denote  the  anomalies   of   $U(1)$'s   whose
coefficients  are  also 1.  Thus the current $J_\mu ^X$  has  the
same anomaly structure as that of the current corresponding to
the model-independent axion.  Therefore, the anomalous $U(1)_X$
gauge boson obtains a mass, absorbing the model-independent axion
as the longitudinal degree.  Then,
{\it there results a global symmetry
below the $U(1)_X$ gauge boson mass which is a PQ symmetry.}
To solve the $\mu$ problem,
we identify $\Lambda_{SU(5)}$ as $M_I$.  $S_1$ and $S_2$, which
are $SU(5)$ quintet and antiquintet, can condense to give
$\langle S_1S_2\rangle\sim M_I^2$, which leads to the desired
magnitude for the $\mu$ term.  But the QCD $\theta$ problem is
not solved automatically
because there is only one available global symmetry.  However,
the possibility of the phase field of the hidden sector
gaugino condensation providing another needed phase has been
argued in Ref. [8].

\noindent 5. {CONCLUSION}

The PQ symmetry has been intruduced to solve the strong CP
problem and the $\mu$ problem in supergravity models.  At the
supergravity level it works.  Superstring extension

of this idea also solves these two problems if there is no
additional confining group except those of the standard model;
but the mechanism for supersymmetry breaking is not clear.
On the other hand, if there exists an additional confining
group for supersymmetry breaking by gaugino condensation,
the $\mu$ problem is solved; but the solution of the strong CP
problem needs an extra phase field.

\noindent {\bf References}

\pointbegin J. E. Kim and H. P. Nilles,
Phys. Lett. B138 (1984) 150.

\point E. Witten, B149 (1984) 351.

\point K. Choi and J. E. Kim, Phys. Lett. B154  (1985) 393.

\point M. Dine, N. Seiberg and E. Witten, Nucl. Phys. B289,
(1987) 589; J. J. Atick, L. J. Dixon and A. Sen, Nucl. Phys.
B292 (1987) 109.

\point
J. E. Kim, Phys. Lett. B207 (1988) 434.

\point 't Hooft, Nucl. Phys. B35 (1971) 167.

\point J.-P. Derengdinger, L. E. Iba$\tilde {\rm n}$ez and H. P.
Nilles, Phys. Lett. B155 (1985) 65; M. Dine, R. Rhom, N. Seiberg,
and E. Witten, Phys. Lett. B156 (1985) 55.

\point
E. J. Chun, J. E. Kim and H. P. Nilles, Nucl. Phys. B370
(1992) 105.

\bye